\documentclass[aps,twocolumn,prl,superscriptaddress,longbibliography,reprint]{revtex4-2}
\usepackage{amsfonts}
\usepackage{mathrsfs}
\usepackage{amsmath}
\usepackage{color}
\usepackage{graphicx}
\usepackage{bm}
\usepackage{amssymb}
\usepackage{xspace}
\usepackage{epstopdf}
\usepackage{longtable}
\usepackage{multirow}
\usepackage{booktabs}
\usepackage{textcomp}
\usepackage[title]{appendix}
\usepackage{float}
\usepackage{dsfont}
\usepackage[colorlinks=true, letterpaper=true, pdfstartview=FitV, linkcolor=blue, citecolor=blue, urlcolor=blue]{hyperref}
\usepackage[]{changes}

\providecommand{\U}[1]{\protect\rule{.1in}{.1in}}

\begin{document}

\title{Intrinsic Dynamic Generation of Spin Polarization by Time-Varying Electric Field}

\begin{abstract}
Electric control of spin in insulators is desired for low-consumption and ultrafast spintronics, but the underlying mechanism remains largely unexplored. Here, we propose an intrinsic effect of dynamic spin generation driven by time-varying electric field. In the intraband response regime, it can be nicely formulated as a Berry curvature effect and  leads to two phenomena that are forbidden in the $dc$ limit: linear spin generation in nonmagnetic insulators and intrinsic N{\'e}el spin-orbit torque in $\mathcal{PT}$-symmetric antiferromagnetic insulators. These phenomena are driven by the time derivative of field rather than the field itself, and have a quantum origin in the first-order dynamic anomalous spin polarizability. Combined with first-principles calculations, we predict sizable effects driven by terahertz field in nonmagnetic monolayer Bi and in antiferromagnetic even-layer MnBi$_2$Te$_4$, which can be detected in experiment.
\end{abstract}

\author{Xukun Feng}
\thanks{These authors contributed equally to this work.}
\affiliation{Science, Mathematics and Technology, Singapore University of Technology and Design, Singapore 487372, Singapore}

\author{Jin Cao}
\thanks{These authors contributed equally to this work.}
\affiliation{Institute of Applied Physics and Materials Engineering, University of Macau, Macau SAR, China}

\author{Zhi-Fan Zhang}
\thanks{These authors contributed equally to this work.}
\affiliation{Interdisciplinary Center for Theoretical Physics and Information Sciences (ICTPIS), Fudan University, Shanghai 200433, China}
\affiliation{Institute for Nanoelectronic Devices and Quantum Computing and State Key Laboratory of Surface Physics, Fudan University, Shanghai 200433, China}

\author{Lay Kee Ang}
\affiliation{Science, Mathematics and Technology, Singapore University of Technology and Design, Singapore 487372, Singapore}

\author{Shen Lai}
\affiliation{Institute of Applied Physics and Materials Engineering, University of Macau, Macau SAR, China}

\author{Hua Jiang}
\email{jianghuaphy@fudan.edu.cn}
\affiliation{Interdisciplinary Center for Theoretical Physics and Information Sciences (ICTPIS), Fudan University, Shanghai 200433, China}
\affiliation{Institute for Nanoelectronic Devices and Quantum Computing and State Key Laboratory of Surface Physics, Fudan University, Shanghai 200433, China}

\author{Cong Xiao}
\email{xiaoziche@gmail.com}
\affiliation{Institute of Applied Physics and Materials Engineering, University of Macau, Macau SAR, China}

\author{Shengyuan A. Yang}
\affiliation{Institute of Applied Physics and Materials Engineering, University of Macau, Macau SAR, China}

\maketitle

Achieving efficient control of spin degree of freedom by electrical means is a central task in spintronics \cite{Review2009,Manchon2019,Fert2024}. One basic mechanism is the so-called electrical spin generation, where a spin polarization $\Delta \bm S$ of electrons is induced by an electric field \cite{Pikus1978,Aronov1989,Edelstein,Kato2004,Stern2006,Culcer2007,Manchon2008,Garate2009,Geller2009,Miron2010,Ohno2010,Miron2011,Vyborny2011}. Recently, motivated by the prospect of ultrafast spintronic devices, there were intensive experimental studies of
spin generation driven by rapidly varying $E$ field, with the time scale down to pico-second (Terahertz) range \cite{yang2017ultrafast,Jungwirth2018,Baltz2018Antiferromagnetic,zelezny2018Review,Gorchon2020,avci2020picosecond,Polley2023picosecond,Kampfrath2023terahertz}.

The existing theories of electrical spin responses are mainly focused on the \emph{dc} limit \cite{Manchon2019}, which may miss important
mechanisms due to temporal variation of $E$ field. For example, there may exist $\Delta \bm S$ induced by
$\partial \tilde{\bm E}/\partial t$, i.e., the rate of change of a time-varying field $\tilde{\bm E}$. This contribution can be significant, since its symmetry under time reversal is distinct from that of \emph{dc} electrical spin generation, thereby, it could potentially lead to qualitatively new physical phenomena. Nevertheless, the symmetry properties and the consequences of such contributions have not been clarified yet.

In this work, we explore this mechanism of dynamic electrical spin generation (DESG). Our focus is on the intraband response regime, where the driving frequency $\omega$ is lower than that required for interband transitions such that intraband processes dominate the response. We show that, in this regime, DESG can be nicely formulated as a Berry curvature effect and permits \emph{intrinsic} contributions which are independent of scattering.
The response tensor of intrinsic DESG can be expressed as a sum of band geometric quantity, the dynamic anomalous spin polarizability (DASP), over occupied states.
Particularly, we uncover that owing to its distinct symmetry character, the intrinsic effect linear in  $\partial \tilde{\bm E}/\partial t$ enables two phenomena that are forbidden in the \emph{dc} limit: (i) linear spin generation in
nonmagnetic insulators; and (ii) intrinsic N\'{e}el spin-orbit torque in spacetime-inversion ($\mathcal{PT}$) symmetric
antiferromagnetic insulators. Combining our theory with first-principles calculations, we predict sizable
effects under THz driving in nonmagnetic 2D Bi and in antiferromagnetic even-layer MnBi$_2$Te$_4$, which can be detected in experiment. Our finding offers a new route for manipulating spins, especially in insulators,
which is promising for low-dissipation ultrafast spintronics.

\textcolor{blue}{\textit{DESG as a Berry curvature effect.}}
In the intraband response regime, the spin polarization induced by a time-varying $E$ field can be well described in the framework of semiclassical theory \cite{Xiao2010}. Following the general approach in Ref.~\cite{Dong2020}, to evaluate spin polarization, we introduce an auxiliary Zeeman-like field $\bm h$, which couples to the spin operator $\hat{\bm s}$ and adds a term $\bm h\cdot \hat{\bm s}$ to the Hamiltonian. Then, to first order in $E$ field,
the spin polarization of an electron wave packet centered at Bloch state $|u_{n\bm k}\rangle$ can be expressed as ($e<0$ is the electron charge)
\begin{equation}
    \bm s_{n\bm k} = \partial_{\bm h} \varepsilon_{n} -e\Omega_{\bm h \bm k,n} \cdot \Tilde{\bm E} -\hbar \Omega_{\bm h t,n},
    \label{Berry}
\end{equation}
where $\varepsilon_n$ is the band energy, $\Omega$'s are Berry curvature tensors for band $n$ in $h$-$k$ and $h$-$t$ spaces,
and it is understood that the limit $h\rightarrow 0$ is taken at the end of evaluation.

The first term on the right hand side of (\ref{Berry}) is the spin polarization in the absence of $E$ field, and the remaining two terms give the field-induced spin polarization. The second term is a contribution proportional to the field itself, with the Berry curvature tensor \cite{Franz2010,Kurebayashi2014,Freimuth2014,Yuriy2017,Dong2020,Cheng2022,Xiao2022NLSOT,Cheng2024}
\begin{equation}
   \Omega_{\bm h \bm k,n}= \frac{2}{\hbar}\sum_{m \neq n} \frac {\operatorname{Re}(\bm{s}_{nm} \bm{\mathcal{A}}_{mn})}{\omega_{nm}},
   \label{ASP}
\end{equation}
where $\bm s_{nm}=\langle u_n|\hat{\bm s}|u_m\rangle$ is the spin matrix element, $\bm{\mathcal{A}}_{mn}=\langle u_m|i\partial_{\bm k}|u_n\rangle$ is the interband $k$-space Berry connection, and $\hbar \omega_{nm} \equiv \varepsilon_n-\varepsilon_m$.


\begin{table*}
\caption{\label{tabi}Comparison of driving mechanisms for electrical spin generation in nonmagnetic (NM) and antiferromagnetic (AFM) systems. Here, $t$ means a translation between spin sublattice \cite{Zelezny2014,Zelezny2017,Shao2020NAHE}. $\checkmark$ ($\times$) indicates a nonequilibrium spin density can (cannot) be induced in corresponding systems.}
\begin{ruledtabular}
\begin{centering}
\begin{tabular}{ccccc}
Mechanism & \begin{tabular}[c]{@{}c@{}}Effective in\\NM insulators\end{tabular} & \begin{tabular}[c]{@{}c@{}}Effective in\\AFM insulators\end{tabular} & \begin{tabular}[c]{@{}c@{}}$\Delta\boldsymbol{S}$ in\\$t\mathcal{T}$-Symmetric AFMs\end{tabular} & \begin{tabular}[c]{@{}c@{}}$\Delta\boldsymbol{S}$ in\\$\mathcal{PT}$-Symmetric AFMs\end{tabular}\tabularnewline
\hline
Edelstein effect & $\times$ & \multicolumn{1}{c}{$\times$} & \multirow{2}{*}{Non-staggered} & \multirow{2}{*}{Staggered}\tabularnewline
$\Upsilon^{\left(1\right)}$ [Eq.~(\ref{dynamical ASP-1})] & $\checkmark$ & $\checkmark$ &  & \tabularnewline
\cline{1-3}
$\Upsilon^{\left(0\right)}$ [Eq.~(\ref{ASP})] & $\times$ & \multicolumn{1}{c}{$\checkmark$} & \multirow{2}{*}{Staggered} & \multirow{2}{*}{Non-staggered}\tabularnewline
$\Upsilon^{\left(2\right)}$ [Eq.~(\ref{dynamical ASP-2})] & $\times$ & $\checkmark$ &  & \tabularnewline
\end{tabular}
\par\end{centering}
\end{ruledtabular}
\end{table*}


On the other hand, the last term with $\Omega_{\bm h t}$ manifests qualitatively different physics, as it captures the DESG responses
due to temporal variation of $E$ field. To obtain an explicit expression, let us take a harmonic field $\Tilde{\bm E}(t)= \bm E \cos \omega t$. Then, the first order correction to state $|u_n\rangle$ is given by
\begin{equation}
    |\delta u_n\rangle = -\frac{e}{2\hbar}\sum_{m \neq n}\bm{E}\cdot \bm{\mathcal{A}}_{mn}\bigg(\frac{e^{-i\omega t}}{\omega_{nm}+\omega}+\frac{e^{i\omega t}}{\omega_{nm}-\omega}\bigg)|u_m\rangle.
    \label{perturbation}
\end{equation}
Up to this order, one can easily verify that the Berry connection with respect to time is zero, so we have
\begin{equation}\label{omegaht}
    \Omega_{\bm h t,n}= -\partial_t \mathfrak{\bm a}_n,
\end{equation}
where $\mathfrak{\bm a}$ is the Berry connection with respect to $\bm h$. Direct calculation gives the following gauge invariant expression:
\begin{equation}
    \mathfrak{\bm a}_n = -\frac{2e}{\hbar^2}\operatorname{Im}\sum_{m \neq n} \frac{\bm{s}_{nm} \bm{\mathcal{A}}_{mn}}{\omega_{nm}^2-\omega^2} \cdot \bigg(\Tilde{\bm E} -\frac{i\partial_t \Tilde{\bm E}}{\omega_{nm}}\bigg).
    \label{vector potential}
\end{equation}
This $\mathfrak{\bm a}$ can be interpreted as a vector potential in $h$ space, then Eq.~(\ref{omegaht}) suggests that
$\Omega_{\bm h t}$ represents a $h$-space electric field, which is induced by the temporary variation of real-space electric field. From Eq.~(\ref{vector potential}), there are two contributions to $\Omega_{\bm h t}$, one depends on $\partial_t
\Tilde{\bm E}$, and the other on $\partial_t^2
\Tilde{\bm E}$.

Summarizing the above results, the field-induced spin polarization $\delta\bm s$ for a wave packet can be expressed as
\begin{align}
     \delta \bm s_{n \bm k} = \Upsilon^{(0)}_n \cdot \Tilde{\bm E}
    +\Upsilon^{(1)}_n \cdot \partial_t \Tilde{\bm E}
    +\Upsilon^{(2)}_n \cdot \partial^2_t \Tilde{\bm E}
    \label{Berry-1},
\end{align}
where $\Upsilon^{(0)}_n=-e\Omega_{\bm h \bm k,n}$ is also known as the anomalous spin polarizability \cite{Xiao2023NLSOT},
the two second-rank tensors
\begin{align}
     \Upsilon^{(1)}_n=
     -\frac{2e}{\hbar}\sum_{m \neq n} \frac {\operatorname{Im}(\bm{s}_{nm} \bm{\mathcal{A}}_{mn})}{\omega_{nm}^2-\omega^2}
     \label{dynamical ASP-1}
\end{align}
and
\begin{align}
      \Upsilon^{(2)}_n=
    \frac{2e}{\hbar}\sum_{m \neq n} \frac {\operatorname{Re}(\bm{s}_{nm} \bm{\mathcal{A}}_{mn})}{\omega_{nm} (\omega_{nm}^2-\omega^2)}
    \label{dynamical ASP-2}
\end{align}
are the first-order and second-order DASP, respectively.

The total induced spin polarization is obtained by summing over all occupied states:
$
    \Delta \bm S = \int [d\boldsymbol{k}] f_{n\bm k} \delta \bm s_{n \bm k}
$, where $f$ is the distribution function and $[d\boldsymbol{k}]$ is shorthand notation for $\sum_{n} d\boldsymbol{k}/(2\pi)^{d}$ with $d$ the dimension of the system. As mentioned, here, we are most interested in the
intrinsic response, with $f$ given by the equilibrium Fermi distribution $f_0$. Intrinsic responses are of great importance in condensed matter research, as they represent properties intrinsic to each material system. In addition, the intrinsic DESG studied here will be dominating in insulating systems, as we will discuss below. Separating contributions according to the order of time derivative (as in Eq.~(\ref{Berry-1})), we obtain
\begin{equation}\label{deltaS}
  \Delta \bm S =\sum_i  \Delta^{(i)} \bm S,
\end{equation}
where $i=0, 1, 2$, and
\begin{align}
    \Delta^{(i)} \bm S = \int [d\boldsymbol{k}] f_0 \Upsilon^{(i)} \cdot \partial^i_t \Tilde{\bm E}.
    \label{response}
\end{align}
Here, the $i=0$ contribution reproduces the previous result for \emph{dc} electrical spin generation \cite{Franz2010,Kurebayashi2014,Freimuth2014,Yuriy2017,Dong2020,Cheng2022,Xiao2022NLSOT,Cheng2024}, the other two contributions $i=1,2$ give the intrinsic DESG. The response tensors for DESG are given by the integrals of DASPs over the occupied states.

We have a few remarks before proceeding. First of all, as we noted, compared with \emph{dc} response,
the DESG response $\Delta^{(1)} S$ has distinct symmetry character under time reversal operation: it is time-reversal-even ($\mathcal{T}$-even). In contrast,  $\Delta^{(2)} S$, like  $\Delta^{(0)} S$, is $\mathcal{T}$-odd.
This means $\Delta^{(0)} S$ and $\Delta^{(2)} S$ appear only in magnetic systems, whereas $\Delta^{(1)} S$ occurs in both magnetic and nonmagnetic systems. In other words, of the three, only $\Delta^{(1)} S$ survives in nonmagnetic systems.

As a result of its distinct symmetry, the DESG $\Delta^{(1)} S$ can lead to unprecedented phenomena that are impossible in \emph{dc} limit. A prominent example is the linear spin generation in nonmagnetic insulators. Nonmagnetic materials can only host $\mathcal{T}$-even response, which, under $dc$ driving, must involve carrier scattering and requires the presence of Fermi surface, i.e., a metallic state (this is also known as the Edelstein effect) \cite{Pikus1978,Aronov1989,Edelstein}. In contrast, the first-order DASP $\Upsilon^{(1)}$ enables scattering-independent $\mathcal{T}$-even spin response, which does not require a Fermi surface. In nonmagentic insulators, this intrinsic DESG $\Delta^{(1)} S$
would be the dominant effect, as extrinsic effects from scattering are suppressed by the insulating gap.
Furthermore, in $\mathcal{PT}$-symmetric antiferromagnets, a research focus is to generate N{\'e}el spin-orbit torque \cite{Manchon2019}. It is a field-like torque arising from $\mathcal{T}$-even spin response that is staggered on opposite magnetic sublattices \cite{Zelezny2017}. In $dc$ limit, it can only stem from a staggered Edelstein effect \cite{Zelezny2014,Zelezny2017}, which is an extrinsic Fermi surface response thus does not work in insulators. The DESG $\Delta^{(1)} S$ naturally provides a solution to this problem. By symmetry, it automatically gives a N\'{e}el torque
in $\mathcal{PT}$ antiferromagnetic insulators, and this torque is of intrinsic nature. A comparison of the different mechanisms is summarized in Table.~\ref{tabi}.
In addition, since the spin polarization and the torque generated here are in the insulating state, the dissipation via Joule heating can be minimized. This is a great advantage for spintronics applications.

Back to Eq.~(\ref{response}), we note that although the sum is over all occupied bands, the main contribution is actually from the
low-energy conduction and valence bands close to chemical potential $\mu$. This is because by substituting the expressions of $\Upsilon^{(i)}$ into Eq.~(\ref{response}), the terms involving two occupied bands $n$ and $m$ must cancel exactly after summation. Only those terms with $n$ occupied and $m$ unoccupied survive, and such terms are large when bands $m$ and $n$ are close to $\mu$.


Finally, we mention that our result in Eqs.~(\ref{deltaS}-\ref{response}) are fully consistent with that obtained by  standard linear response theory \cite{Sipe2022spin,supp}, justifying its validity. However, in linear response theory, the three contributions are entangled in a single expression. It is not straightforward to separate them and to clarify their symmetry characters. In comparison, our approach here has the advantage that it clearly separates the different contributions, and it is also physically appealing by associating the responses to band geometric properties (Berry curvatures).

\begin{figure}[t!]
	\centering
	\includegraphics[width=8cm]{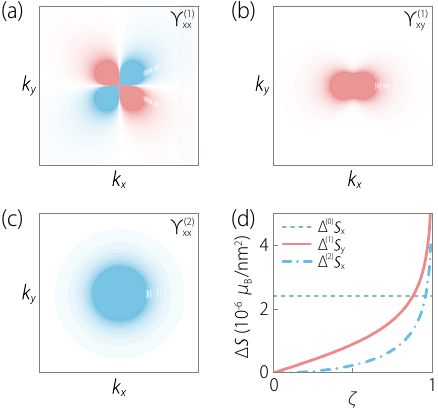}
	\caption{DESG for model (\ref{model}). (a-c) The $k$-space distribution of (a) $\Upsilon^{(1)}_{xx}$, (b) $\Upsilon^{(1)}_{xy}$, and (c) $\Upsilon^{(2)}_{xx}$ of the lower Dirac band. (d) Amplitudes of generated spin polarizations due to the three mechanisms versus $\zeta$. In the calculation, $v=5 \times 10^5$ m/s, $\Delta = 0.15$ eV, and $E_x =10^4 $~V/m. In (a-c), we take $\zeta\equiv\hbar\omega/(2\Delta)=0.5$.}
	\label{fig1}
\end{figure}

\textcolor{blue}{\textit{Gapped Dirac model.}}
To illustrate features of DASP and the resulting DESG, we first apply our theory to the 2D gapped Dirac model, which may describe the surface of a topological insulator in contact with a ferromagnet:
\begin{equation}
    H= \hbar v \bm \sigma \cdot (\bm k \times \hat{\bm z})+\Delta\sigma_z.
    \label{model}
\end{equation}
Here, $\bm \sigma$ is the vector of Pauli matrices for spin, $v$ and $\Delta$ are taken as real positive parameters.

This model breaks time reversal symmetry, so all three contributions in (\ref{response}) exist. The intrinsic \emph{dc} spin response (corresponding to $\Delta^{(0)}S$) of this model has been investigated before~\cite{Franz2010}. We shall focus on the dynamic responses. The DASPs take the form of
\begin{align}
\Upsilon^{(1)} &=\frac{\hbar ^5v^2}{\varepsilon ^3\left( 4\varepsilon ^2-\hbar ^2\omega ^2 \right)}\left[ \begin{matrix}
	-k_xk_y&	 k_{x}^{2}+\frac{\Delta ^2}{\hbar ^2v^2}\\
	-k_{x}^{2}-\frac{\Delta ^2}{\hbar ^2v^2}&		k_xk_y\\
\end{matrix} \right],\nonumber
\\
\Upsilon^{(2)} &=\frac{-\hbar^5v\Delta}{2\varepsilon ^3\left( 4\varepsilon ^2-\hbar ^2\omega ^{{2}} \right)}\left[ \begin{matrix}
	1&		0\\
	0&		1\\
\end{matrix} \right],
\end{align}
with $\varepsilon = \sqrt{\hbar^2v^2k^2+\Delta^2}$. Their distributions in $k$ space are plotted in Fig.~\ref{fig1}(a-c). One observes that they are concentrated around the minimal gap region.
Take $\tilde{\bm E}(t)=E_x \cos (\omega t) \hat{x}$ and assume
$\mu$ lies in the band gap. The spin responses are found to be $\Delta^{(0)}\bm S =S_0 \cos (\omega t)\hat{x}$,
\begin{align}
\Delta^{(1)}\bm S &=S_0\sin \omega t\left( \frac{1+\zeta^{-2}}{2}\mathrm{arcoth}\frac{1}{\zeta}-\frac{1}{2\zeta} \right)\hat{y}, \nonumber
\\
\Delta^{(2)}\bm S &=S_0\cos \omega t\left( \frac{1}{\zeta}\mathrm{arcoth}\frac{1}{\zeta}-1 \right)\hat{x},
\end{align}
where $S_0 = eE_x/4\pi v$ and $\zeta\equiv\hbar
\omega/(2\Delta)\in (0,1)$. One observes that the $\mathcal{T}$-even response $\Delta^{(1)}\bm S$ may have a direction different from the other two contributions as well as the applied field.
Figure~\ref{fig1}(d) plots their amplitudes versus $\zeta$. One finds that both DESG contributions increase with frequency, and large values can be obtained when $\hbar\omega$ is close to the band gap, i.e., the upper bound of the considered regime.


\begin{figure}[t!]
	\includegraphics[width=8cm]{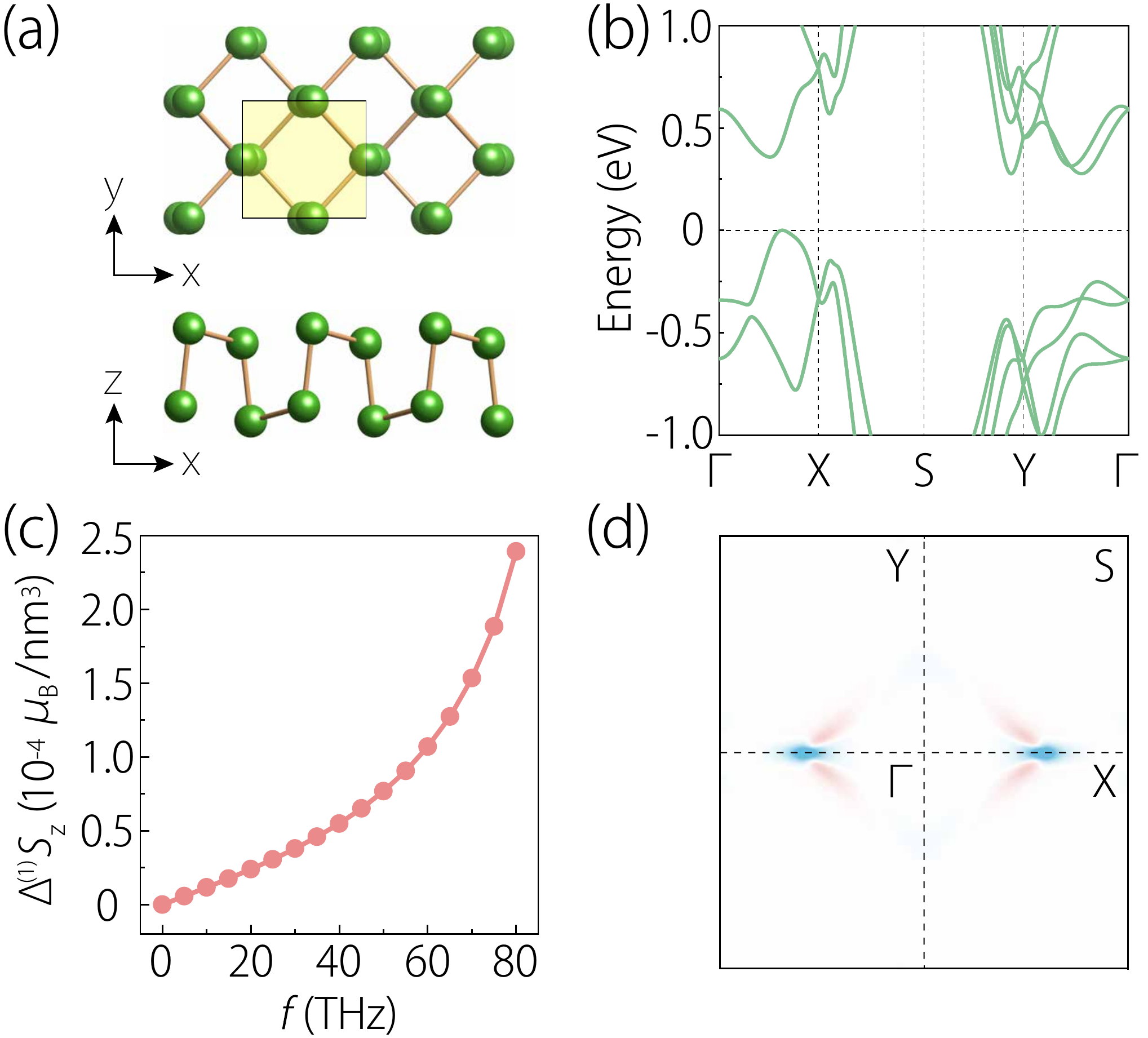}
	\caption{(a) Top and side views of monolayer Bi. The shaded rectangle indicates the unit cell. (b) Calculated band structure for monolayer Bi. (c) Amplitude of DESG $\Delta^{(1)} S_z$ versus driving frequency. (d) $k$-space distribution of $\Upsilon^{(1)}_{zy}$ of valence bands at $50$ THz. 
		\label{fig2}}
\end{figure}

\textcolor{blue}{\textit{Application to monolayer Bi.}} Next, we combine our theory with first-principles calculations to study DESG in a nonmagnetic insulator, the monolayer Bi. Previous studies have identified 2D Bi with black phosphorus structure as an elemental ferroelectric, with broken inversion symmetry~\cite{xiao2018elemental, gou2023two}, which permits the linear spin generation.

The lattice structure of monolayer Bi is depicted in Fig.~\ref{fig2}(a), which has $C_{2v}$ point group. The calculated band structure in Fig.~\ref{fig2}(b) displays a global gap $\sim 0.280$ eV (calculation details in Ref.~\cite{supp}), which is consistent with recent experimental observation~\cite{gou2023two}. The minimal local gap $\sim 0.394$ eV is located along the $\Gamma$-X path. Such a gap size corresponds to an upper bound $\sim 100$ THz for intraband response regime.

Symmetry of this system allows a nonzero $\Delta^{(1)} S_z$ generated by a time-varying field along the $y$ direction.
Figure~\ref{fig2}(c) shows the calculated amplitude of $\Delta^{(1)} S_z$ versus driving frequency, at a moderate field strength of $10^7$ V/m \cite{Jungwirth2018}. One observes that at frequency $\geq 10$ THz, a spin polarization greater than $10^{-5}$ $\mu_\text{B}$/nm$^3$ can be achieved. This value is significant, considering that existing techniques can detect spin polarization down to $10^{-9}$ $\mu_\text{B}/\mathrm{nm^{3}}$ \cite{Kato2004,Stern2006}. The evolution of spin polarization can be
probed by time-resolved magneto-optical Kerr measurement, which has achieved a resolution $\sim 100$ fs~\cite{Igarashi2023optically,Polley2023picosecond,Gray2024}. In Fig.~\ref{fig2}(d), we plot the distribution of DASP $\Upsilon^{(1)}_{zy}$ for occupied bands, which reveals hot spots at the minimal local gap regions. These features in Fig.~\ref{fig2}(c) and \ref{fig2}(d) are consistent with those found in the model study.

\begin{figure}[t!]
\includegraphics[width=8cm]{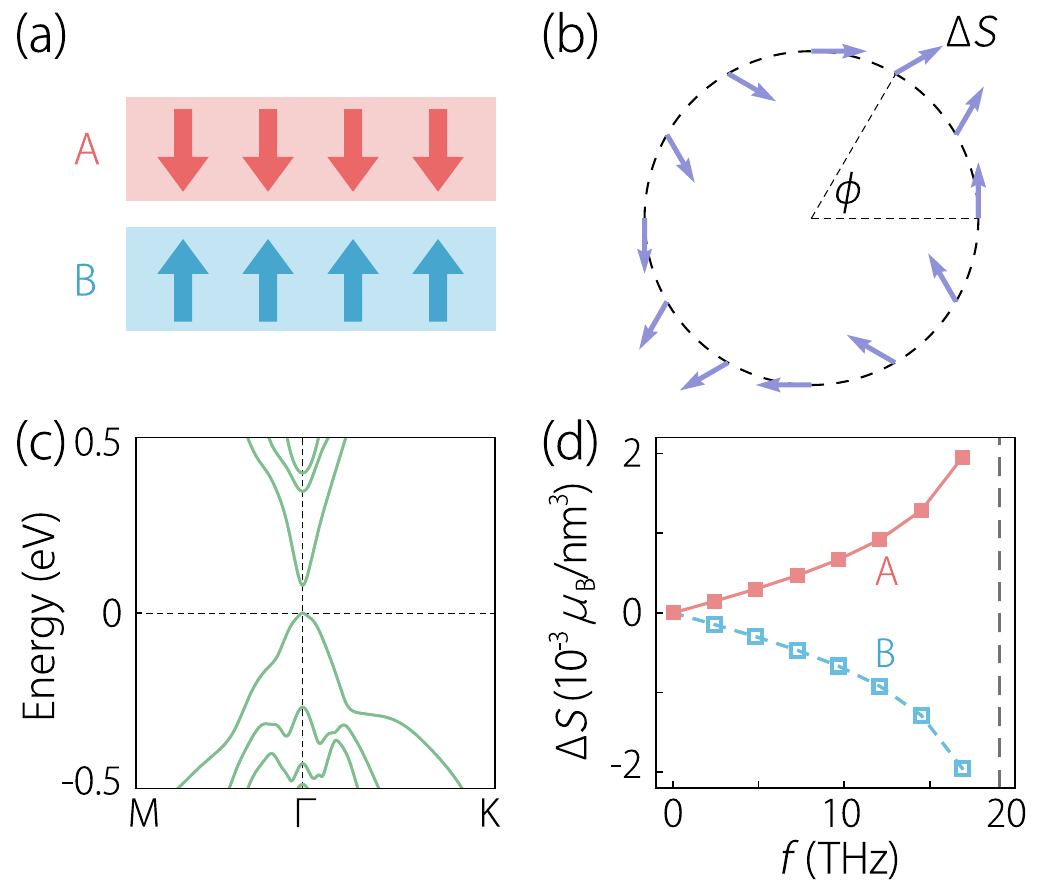}
\caption{\label{fig4}(a) Schematic drawing for bilayer MnBi$_2$Te$_4$, which has antiferromagnetic ground state. The two magnetic sublattices $A$ and $B$ correspond to the two layers. The local moments are in the out-of-plane direction. 
(b) Direction of $\Delta^{(1)} \boldsymbol{S}$ on sublattice $A$ versus the orientation of in-plane driving field (labeled by polar angle $\phi$).
(c) Calculated band structures of bilayer MnBi$_2$Te$_4$. (d) DESG $\Delta^{(1)} S_y$ on the two magnetic sublattices (solid and dashed lines) for field applied along $x$ direction. The insulating gap is indicated by the vertical line.}
\end{figure}

\textcolor{blue}{\textit{Intrinsic N\'{e}el spin-orbit torque in MnBi$_2$Te$_4$.}}
The possibility to utilize antiferromagnets for ultrafast spintronics has been a focus of recent research \cite{Baltz2018Antiferromagnetic,Jungwirth2018,zelezny2018Review,Cheng2016,Kampfrath2023terahertz}.
A central task is the generation of a N\'{e}el torque. As we discussed, DESG provides a new way to solve this problem, especially for $\mathcal{PT}$-symmetric antiferromagnetic insulators, where the conventional method based on the staggered Edelstein effect \cite{wadley2016electrical,Godinho2018} fails.


We demonstrate this in 2D MnBi$_2$Te$_4$ \cite{Gong2019,Chulkov2019,Deng2020,Zhang2019,Xu2019,Otrokov2019,Zeugner2019,Yan2019,Lee2019,Cui2019,Liu2020} with an even numbers of septuple layers, which is an antiferromagnetic semiconductor with out-of-plane magnetic moments (Fig.~\ref{fig4}(a)).
Each septuple layer of MnBi$_2$Te$_4$ has $3m^{\prime }$ magnetic point group symmetry, which allow nonzero DASP $\Upsilon^{(1)}_{xy}=\Upsilon^{(1)}_{yx}$. It follows that under an in-plane time-varying $E$ field,
DESG in one sublattice $A$ is in the direction $ ({\hat{x}}\partial_t \Tilde{E}_y+{\hat{y}}\partial_t \Tilde{E}_x)$, as illustrated in Fig.~\ref{fig4}(c), and it is opposite in the other sublattice $B$.


Consider bilayer MnBi$_2$Te$_4$, where the two sublattices just correspond to the two layers (Fig.~\ref{fig4}(a)). Figure \ref{fig4}(b) shows the calculated band structure. The band gap is $\sim79$~meV, which agrees with previous report~\cite{Chulkov2019}.
To evaluate the sublattice spin polarization, one just needs to project $\delta  s_{n\bm k}$ in (\ref{Berry-1}) onto each layer here.
Under an \emph{ac} field in $x$ direction with amplitude of $10^7$~V/m, the resulting spin-$y$ polarization is shown in Fig.~\ref{fig4}(d). One finds that the spin polarization is indeed staggered on the two sublattices. The effect vanishes in \emph{dc} limit, shows approximately linear dependence at low frequencies, and departs from linearity as frequency approaches the insulating gap (indicated by the vertical line in Fig.~\ref{fig4}(d)). The nonequilibrium spin density on each spin sublattice can reach $2 \times 10^{-3}$~$\mu_B/$nm$^3$ at $17~$THz.
We also find result of a similar magnitude for four-layer MnBi$_2$Te$_4$ (see \cite{supp}).

This DESG gives rise to a staggered effective magnetic field with magnitude $B_{\mathrm{eff}}\sim \left(J_{\mathrm{ex}}/\mu_{\mathrm{B}}\right)\Delta^{(1)}S/M_s$~\cite{Zelezny2017}, where the exchange strength $J_{\mathrm{ex}}\sim 1~$eV \cite{Zelezny2014}, and $M_s\sim20.7~\mu_B/$nm$^3$ is the saturation magnetization of one sublattice.
Taking $\Delta^{(1)}S\sim 2 \times 10^{-3}$~$\mu_B/$nm$^3$,
we find $B_{\mathrm{eff}}$ can reach $\sim1.67$~T, which is orders of magnitude larger than those by Edelstein effect found in antiferromagnetic metals Mn$_2$Au and CuMnAs~\cite{wadley2016electrical,Zelezny2017}. This value is also much larger than the magnetic anisotropy field $\sim$ 225 mT for bilayer MnBi$_2$Te$_4$~\cite{Lujan2022Magnons}, making it promising for the switching of N\'{e}el order.
The induced spin polarization and magnetic dynamics can be detected by techniques such as time-resolved magneto-optical probes \cite{Fan2014,Wang2015MOKE}, spin Hall magneto-resistance \cite{Saitoh2013SMR,brataas2013insulating}, or X-ray magnetic linear dichroism \cite{Dhesi2017XMLD,wadley2018XMLD}.



\textcolor{blue}{\textit{Discussion.}}
We have proposed an intrinsic effect of DESG, clarified its distinct symmetry character, and unveiled its potential for manipulating spin and generating N\'{e}el torque in insulator systems (with the advantage of low heat dissipation), where the conventional mechanisms do not work. Our finding thus greatly broadens the scope of spin source material platforms, which are so far based mainly on
metals.

Like other band geometric quantities, DASPs represent a kind of interband coherence. They are enhanced when the driving frequency is close to the local gap. One may regard that the \emph{ac} driving effectively reduces the gap, which enhances
interband coherence. This physics is in the same spirit as previous proposed optical enhancement of topological transport in semiconductors \cite{Yao2007}.

Finally, intrinsic DESG should widely exist in materials with broken inversion symmetry. The exact form of response tensor is determined by crystal symmetry. For example, in monolayer \textit{T}$_d$-WTe$_2$ \cite{tang2017quantum,xu2018electrically}, we find that a large in-plane spin polarization $\sim 1.7 \times 10^{-5}$ $\mu_{B}$/nm$^3$ can be induced by in-plane field  of $10$ THz and amplitude $10^7$ V/m in the orthogonal direction \cite{supp}. In addition, in certain antiferromagnets with $t\mathcal{T}$ symmetry \cite{Zelezny2014,Zelezny2017,Shao2020NAHE}, where $t$ is fractional translation between magnetic sublattices, as indicated in Table~\ref{tabi}, a N{\'e}el torque can be produced by intrinsic $\Delta^{(2)}S$ along with $\Delta^{(0)}S$, instead of $\Delta^{(1)}S$.



\bibliographystyle{apsrev4-2}
\bibliography{ref}

\end{document}